# Observation of a New Charmed Strange Meson

Y. Kubota,[1] M. Lattery,[1] J.K. Nelson,[1] S. Patton,[1] D. Perticone,[1] R. Poling,[1] V. Savinov,[1] S. Schrenk,[1] R. Wang,[1] M.S. Alam,[2] I.J. Kim,[2] B. Nemati,[2] J.J. O'Neill,[2] H. Severini,[2] C.R. Sun,[2] M.M. Zoeller,[2] G. Crawford,[3] C. M. Daubenmier,[3] R. Fulton,[3] D. Fujino,[3] K.K. Gan,[3] K. Honscheid,[3] H. Kagan,[3] R. Kass,[3] J. Lee,[3] R. Malchow,[3] F. Morrow,[3] Y. Skovpen,[3*] M. Sung,[3] C. White,[3] F. Butler,[4] X. Fu,[4] G. Kalbfleisch,[4] W.R. Ross,[4] P. Skubic,[4] J. Snow,[4] P.L. Wang,[4] M. Wood,[4] D.N. Brown,[5] J.Fast ,[5] R.L. McIlwain,[5] T. Miao,[5] D.H. Miller,[5] M. Modesitt,[5] D. Payne,[5] E.I. Shibata,[5] I.P.J. Shipsey,[5] P.N. Wang,[5] M. Battle,[6] J. Ernst,[6] Y. Kwon,[6] S. Roberts,[6] E.H. Thorndike,[6] C.H. Wang,[6] J. Dominick,[7] M. Lambrecht,[7] S. Sanghera,[7] V. Shelkov,[7] T. Skwarnicki,[7] R. Stroynowski,[7] I. Volobouev,[7] G. Wei,[7] P. Zadorozhny,[7] M. Artuso,[8] M. Goldberg,[8] D. He,[8] N. Horwitz,[8] R. Kennett,[8] R. Mountain,[8] G.C. Moneti,[8] F. Muheim,[8] Y. Mukhin,[8] S. Playfer,[8] Y. Rozen,[8] S. Stone,[8] M. Thulasidas,[8] G. Vasseur,[8] G. Zhu,[8] J. Bartelt,[9] S.E. Csorna,[9] Z. Egyed,[9] V. Jain,[9] K. Kinoshita,[10] K.W. Edwards,[11] M. Ogg,[11] D.I. Britton,[12] E.R.F. Hyatt,[12] D.B. MacFarlane,[12] P.M. Patel,[12] D.S. Akerib,[13] B. Barish,[13] M. Chadha,[13] S. Chan,[13] D.F. Cowen,[13] G. Eigen,[13] J.S. Miller,[13] C. O'Grady,[13] J. Urheim,[13] A.J. Weinstein,[13] D. Acosta,[14] M. Athanas,[14] G. Masek,[14] H.P. Paar,[14] J. Gronberg,[15] R. Kutschke,[15] S. Menary,[15] R.J. Morrison,[15] S. Nakanishi,[15] H.N. Nelson,[15] T.K. Nelson,[15] C. Qiao,[15] J.D. Richman,[15] A. Ryd,[15] H. Tajima,[15] D. Schmidt,[15] D. Sperka,[15] M.S. Witherell,[15] M. Procario,[16] R. Balest,[17] K. Cho,[17] M. Daoudi,[17] W.T. Ford,[17] D.R. Johnson,[17] K. Lingel,[17] M. Lohner,[17] P. Rankin,[17] J.G. Smith,[17] J.P. Alexander,[18] C. Bebek,[18] K. Berkelman,[18] K. Bloom,[18] T.E. Browder,[18] D.G. Cassel,[18] H.A. Cho,[18] D.M. Coffman,[18] P.S. Drell,[18] R. Ehrlich,[18] M. Garcia-Sciveres,[18] B. Geiser,[18] B. Gittelman,[18] S.W. Gray,[18] D.L. Hartill,[18] B.K. Heltsley,[18] C.D. Jones,[18] S.L. Jones,[18] J. Kandaswamy,[18] N. Katayama,[18] P.C. Kim,[18] D.L. Kreinick,[18] G.S. Ludwig,[18] J. Masui,[18] J. Mevissen,[18] N.B. Mistry,[18] C.R. Ng,[18] E. Nordberg,[18] J.R. Patterson,[18] D. Peterson,[18] D. Riley,[18] S. Salman,[18] M. Sapper,[18] F. Würthwein,[18] P. Avery,[19] A. Freyberger,[19] J. Rodriguez,[19] R. Stephens,[19] S. Yang,[19] J. Yelton,[19] D. Cinabro,[20] S. Henderson,[20] T. Liu,[20] M. Saulnier,[20] R. Wilson,[20] H. Yamamoto,[20] T. Bergfeld,[21] B.I. Eisenstein,[21] G. Gollin,[21] B. Ong,[21] M. Palmer,[21] M. Selen,[21] J. J. Thaler,[21] A.J. Sadoff,[22] R. Ammar,[23] S. Ball,[23] P. Baringer,[23] A. Bean,[23] D. Besson,[23] D. Coppage,[23] N. Copty,[23] R. Davis,[23] N. Hancock,[23] M. Kelly,[23] N. Kwak,[23] and H. Lam[23]

(CLEO Collaboration)

[1] *University of Minnesota, Minneapolis, Minnesota 55455*
[2] *State University of New York at Albany, Albany, New York 12222*
[3] *Ohio State University, Columbus, Ohio, 43210*
[4] *University of Oklahoma, Norman, Oklahoma 73019*
[5] *Purdue University, West Lafayette, Indiana 47907*
[6] *University of Rochester, Rochester, New York 14627*
[7] *Southern Methodist University, Dallas, Texas 75275*
[8] *Syracuse University, Syracuse, New York 13244*
[9] *Vanderbilt University, Nashville, Tennessee 37235*
[10] *Virginia Polytechnic Institute and State University, Blacksburg, Virginia, 24061*
[11] *Carleton University, Ottawa, Ontario K1S 5B6 and the Institute of Particle Physics, Canada*
[12] *McGill University, Montréal, Québec H3A 2T8 and the Institute of Particle Physics, Canada*
[13] *California Institute of Technology, Pasadena, California 91125*
[14] *University of California, San Diego, La Jolla, California 92093*
[15] *University of California, Santa Barbara, California 93106*
[16] *Carnegie-Mellon University, Pittsburgh, Pennsylvania 15213*
[17] *University of Colorado, Boulder, Colorado 80309-0390*
[18] *Cornell University, Ithaca, New York 14853*
[19] *University of Florida, Gainesville, Florida 32611*
[20] *Harvard University, Cambridge, Massachusetts 02138*
[21] *University of Illinois, Champaign-Urbana, Illinois, 61801*
[22] *Ithaca College, Ithaca, New York 14850*
[23] *University of Kansas, Lawrence, Kansas 66045*

## Abstract

Using the CLEO-II detector, we have obtained evidence for a new meson decaying to $D^0 K^+$. Its mass is $2573.2^{+1.7}_{-1.6} \pm 0.8 \pm 0.5$ MeV/$c^2$ and its width is $16^{+5}_{-4} \pm 3$ MeV/$c^2$. Although we do not establish its spin and parity, the new meson is consistent with predictions for an $L = 1$, $S = 1$, $J^P = 2^+$ charmed strange state.

PACS numbers: 13.25.+m, 14.40.Jz

Typeset using REVTEX

*Permanent address: INP, Novosibirsk, Russia

Mesons whose quarks have one unit of orbital angular momentum exist in four states. In mesons with one heavy quark, models inspired by Heavy Quark Symmetry indicate we should consider the four states as two doublets [1,2]. The members of one doublet, whose light quark's total angular momentum is $j_\ell = 3/2$ are relatively narrow; they have $J^P = 1^+,\ 2^+$. The mesons in the other doublet, with $j_\ell = 1/2$ and $J^P = 0^+,\ 1^+$, are predicted to be very broad [2]. When the heavy quark is a charm quark, the light quark can be either an up, a down, or a strange quark. Thus there should be 12 $L = 1$ charmed mesons: 6 relatively narrow states and 6 broad states.

All of these narrow resonances have been observed [3,4], except for the charmed strange $J^P = 2^+$ meson, designated $D_{s2}^{*+}$ [5]. The allowed decay modes of the $D_{s2}^{*+}$ are $DK$ and $D^*K$, both proceeding through a $D$-wave. Because of the limited phase space, the latter is highly suppressed. Godfrey and Kokoski predict the partial width for the decay to $DK$ to be 6 to 10 times larger than for $D^*K$ [2]. The decay to $D_s^+\pi^0$ is forbidden by isospin conservation; modes such as $D_s^+\pi\pi$ are OZI-suppressed. Thus we have searched for the decays $D_{s2}^{*+} \to D^0K^+$ and $D^{*0}K^+$. Throughout this paper charge-conjugate reactions are implied.

The data used in this analysis were collected with the CLEO-II detector at the Cornell Electron Storage Ring (CESR). The detector consists of a charged particle tracking system surrounded by time-of-flight (TOF) scintillation counters. These are followed by an electromagnetic calorimeter, which consists of 7800 thallium-doped CsI crystals. The inner detector is operated in a 1.5 T solenoidal magnetic field generated by a superconducting coil. Finally, the magnet coil is surrounded by iron slabs and muon counters. A detailed description of the detector can be found elsewhere [6].

The data were taken at center-of-mass energies equal to the masses of the $\Upsilon(3S)$ and $\Upsilon(4S)$, and in the continuum above and below the $\Upsilon(4S)$. The total integrated luminosity is 2.16 fb$^{-1}$. Events were required to have a minimum of five charged tracks and energy in the calorimeter of at least 15% of the center-of-mass energy.

Specific ionization measurements from the main drift chamber and TOF measurements were used to identify charged particles. Particles were required to pass a consistency cut for the hypothesis in question: kaon or pion. We define $\chi^2 \equiv (\frac{\Delta Q}{\sigma_Q})^2 + (\frac{\Delta T}{\sigma_T})^2$, where $\Delta Q$ is the difference between the measured and expected specific ionization for the hypothesis. Similarly, $\Delta T$ is the difference between the measured and expected time for the same hypothesis. Time-of-flight information was only used when the track's polar angle with respect to the beamline, $\theta$, met the requirement $|\cos\theta| \leq 0.71$. For a track to be considered a pion or kaon candidate, $\chi^2 \leq 6.25$ was required for the corresponding hypothesis.

Energy clusters in the calorimeter not matched to a charged track and which had $E \geq$ 50 MeV were accepted as photon candidates. To reconstruct $\pi^0$'s, we used pairs of photons from the "good barrel" region, $|\cos\theta| \leq 0.71$, which has the best energy resolution, or one photon from the "good barrel" and one photon from the "good endcap" region ($0.86 \leq |\cos\theta| \leq 0.94$), which has nearly as good resolution. The invariant mass of the two photons was required to be within 2.5 standard deviations of the $\pi^0$ mass; the $\pi^0$ candidates were then kinematically fit to the $\pi^0$ mass to improve momentum resolution. The $\pi^0$'s used to reconstruct $D^0$'s were required to have a minimum energy of 300 MeV; those used to reconstruct the decay $D^{*0} \to D^0\pi^0$ were only required to have an energy greater than 150 MeV.

We reconstructed $D^0$'s in the decay modes $D^0 \to K^-\pi^+$ and $D^0 \to K^-\pi^+\pi^0$. In both cases, the $D^0$ candidates were required to have a measured invariant mass within 1.65 standard deviations of the observed $D^0$ mass peak. The r.m.s. mass resolutions are 10 MeV$/c^2$ for the $K^-\pi^+$ mode and 14 MeV$/c^2$ for the $K^-\pi^+\pi^0$ mode. The decay angle, $\alpha_{K^-}$, is defined as the angle between the $K^-$ direction in the $D^0$ rest-frame and the $D^0$ direction in the lab frame. We required $D^0 \to K^-\pi^+$ candidates satisfy $\cos\alpha_{K^-} \leq 0.8$. This requirement is effective in reducing the background because the signal is flat in $\cos\alpha_{K^-}$, while the background peaks at 1.

In order to reduce the combinatoric background in the $D^0 \to K^-\pi^+\pi^0$ mode, a parameter is calculated for each $D^0$ candidate, based on its position in the Dalitz plot. The parameter varies from 0 to 1, and depends on the square of the amplitude for decay to the observed location in the Dalitz plot. The calculation takes into account the three most important two-body decays and the non-resonant three-body decay which feed into this final state [7]. The two-body decays included are $K^-\rho^+$, $K^{*-}\pi^+$ and $K^{*0}\pi^0$. We require that the parameter be greater than or equal to 0.1. The efficiency of this cut was determined from the inclusive $D^0 \to K^-\pi^+\pi^0$ sample.

The $D^0$ candidates were then combined with each positively charged track consistent with being a kaon. In the search for $D_{s2}^{*+} \to D^0K^+$, there are several sources of background to consider. There is combinatorial background from the various combinations of real and "fake" $D^0$'s with real and "fake" $K^+$'s. The "fake" $D^0$'s are incorrectly reconstructed $D^0$ candidates; the "fake" $K^+$'s are misidentified tracks, mostly pions. The background from real $D^0$'s and fake $K^+$'s includes a component from the decay of the $D_J^*(2470)^+$ to $D^0\pi^+$. If these pions are misidentified as kaons, the $D^0K^+$ mass reconstructed is "reflected" into the mass region near our expected signal. This contribution to the background has been measured by recalculating the energy of the $K^+$ tracks using the pion mass and the measured momenta. The new momentum-energy four-vectors were then combined with our $D^0$ candidates. We observed a peak near 2470 MeV$/c^2$ in the $D^0\pi^+$ mass distribution. Fitting the peak, we found $27 \pm 21$ events in the $K^-\pi^+$ mode and $23 \pm 16$ events in the $K^-\pi^+\pi^0$ mode. Using a Monte Carlo simulation, we have parameterized a shape for this reflected $D_J^*(2470)^+$ background which will be included in our fits to the data. No other resonance was observed in the $D^0\pi^+$ mass distribution; in particular there was no peak from partially reconstructed $D_J(2440)^+$'s. The widths of the $j_\ell = 1/2$ mesons are predicted to be very large [2], and thus should not significantly modify the shape of the background.

To reduce the background from misidentified pions, we imposed an additional ID requirement on the $K^+$ track. We required that the $\chi^2$ for the pion hypothesis for this track be at least 2 units larger than that for the kaon hypothesis. The effectiveness of this cut was evaluated using the $D_{s1}(2536)^+$ feed-down peak described below. The cut has an efficiency of $79 \pm 10\%$, while reducing the background by about a factor of three.

The decay angle, $\alpha_{K^+}$, is defined as the angle between the direction of the $K^+$ in the $D^0K^+$ rest-frame and the $D^0K^+$ direction in the lab frame. We required that the $D^0K^+$ combinations have $\cos\alpha_{K^+} \leq 0.8$. This reduces the combinatoric background which peaks near $\cos\alpha_{K^+} = 1$, and also eliminates some of the background from the $D_J^*(2470)^+$. The distribution of the signal events in this angle is unknown, except that it must be symmetric about $\cos\alpha_{K^+} = 0$. Our overall efficiency for reconstructing the $D^0K^+$, $D^0 \to K^-\pi^+$ combinations is $29 \pm 4\%$. The efficiency for reconstructing the $D^0K^+$, $D^0 \to K^-\pi^+\pi^0$ combinations is $8.9 \pm 1.4\%$.

Finally, to reduce the background, we take advantage of the hard fragmentation of continuum charm and impose a cut on the $x$ of each $D^0K^+$ combination. We define $x \equiv p/p_m$ and $p_m \equiv \{E_{beam}^2 - [M(D^0K^+)]^2\}^{1/2}$. For the $K^-\pi^+$ mode, we require $x \geq 0.7$; for the $K^-\pi^+\pi^0$ mode we require $x \geq 0.8$ because of the larger combinatoric background. The $D^0K^+$ combinations passing all of the above criteria are shown in Fig. 1. To improve the mass resolution, we calculated the "corrected" mass, $M^* \equiv M(D^0K^+) - M(D^0) + 1864.5$ MeV/$c^2$, using the measured invariant masses and the known $D^0$ mass.

Two features are prominent in Fig. 1: a feed-down peak from the $D_{s1}(2536)^+$ at about 2392 MeV/$c^2$, and a broader peak near 2575 MeV/$c^2$, which is a new resonance. Each of these is discussed below.

The feed-down peak at 2392 MeV/$c^2$ is from the $D_{s1}(2536)^+$, the narrow $c\bar{s}$ $1^+$ state. It decays predominantly to $D^*K$. The $Q$ value for the $D^{*0} \to D^0\pi^0$ decay is very small, so even though the $\pi^0$ is not detected, the peak at 2392 MeV/$c^2$ is still very narrow.

Also shown in Fig. 1 is a histogram of $M^*$ for $(K^-\pi^+)K^+$ and $(K^-\pi^+\pi^0)K^+$ combinations, where the $K^-\pi^+$ and $K^-\pi^+\pi^0$ combinations were chosen from the $D^0$ mass sidebands. The $K^-\pi^+$ combinations were chosen from 1800.0 to 1816.5 MeV/$c^2$ and 1912.5 to 1929.0 MeV/$c^2$. The $K^-\pi^+\pi^0$ combinations were chosen from 1777.0 to 1800.0 MeV/$c^2$ and 1927.0 to 1950.0 MeV/$c^2$. This histogram suggests that about half of our background comes from fake $D^0$'s; the other half must come from real $D^0$'s combined with real and fake $K^+$'s. There appears to be some signal in the sideband histogram under both the feed-down peak and the new resonance. This is due to $D^0 \to K^-\pi^+\pi^0$ events in which a poorly measured photon, or the wrong photon, was used to form the $\pi^0$. We have corrected for this effect in the cross-section calculation below. The mass and width measurements are not significantly influenced by this effect. We have also examined the wrong-sign combinations, $D^0K^-$, and see no enhancements in the invariant mass distribution of such combinations.

To extract the mass and width of the new resonance, we fit the $M^*$ distribution for the $D^0K^+$ combinations, as shown in Fig. 2. The fits were done separately for the two $D^0$ decay modes. To parameterize the signal we used a spin-2 relativisitic Breit-Wigner convoluted with a Gaussian of fixed resolution. The r.m.s resolution, $\sigma$, was determined by a Monte Carlo simulation.

For the $D^0 \to K^-\pi^+$ events the Gaussian had $\sigma = 3.2$ MeV/$c^2$. The background, in the mass range from 2430 to 2750 MeV/$c^2$, was fit with a first-order polynomial, plus the $D_J^*(2470)^+$ background function with a fixed area of 27 events. The fit finds $116^{+30}_{-26}$ signal events. The mass is $2573.3^{+2.2}_{-2.1}$ MeV/$c^2$ and the natural width is $15.6^{+6.5}_{-4.8}$ MeV/$c^2$ (statistical errors only). We fit the $D^0 \to K^-\pi^+\pi^0$ data using the same fitting functions, but with $\sigma = 3.7$ MeV/$c^2$. The number of $D_J^*(2470)^+$ background events was fixed at 23. We find $101^{+29}_{-23}$ signal events. Using this mode, the mass is measured to be $2573.1^{+2.7}_{-2.3}$ MeV/$c^2$, and the natural width to be $17.6^{+9.0}_{-6.0}$ MeV/$c^2$ (statistical errors only), in good agreement with the first mode.

Although the fits were done separately for the two $D^0$ decay modes, the data are combined in Fig. 2 with the sum of the fitting functions. The complete signal and background fit is shown by the solid line. The total background is shown by the dashed line. The dotted line shows the polynomial representing the combinatoric background. The shape of the $D_J^*(2470)^+$ background, with the area scaled up by a factor of 5, is shown at the bottom by the dash-dot line. The total signal has a significance of more than six standard deviations.

Following the nomenclature of the Particle Data Group for a meson of unknown spin [5], we will use the temporary designation "$D^*_{sJ}(2573)^+$" for this new resonance. We estimate the systematic error on the $D^{*+}_{sJ}$–$D^0$ mass difference to be $\pm 0.8$ MeV/$c^2$ and on the width to be $\pm 3$ MeV/$c^2$. This includes contributions from varying the assumed mass resolution and spin, the mass of the $D^*_J(2470)^+$ and the number of events it contributes to the background, the order of the polynomial used for the combinatoric background, the binning, and other details of the fitting. The largest contribution to the systematic error on the width, $\pm 1.8$ MeV/$c^2$, comes from varying the order of the polynomial used to fit the background. The next largest, $\pm 1.5$ MeV/$c^2$, comes from the uncertainty in the number of $D^*_J(2470)^+$ events in the background. The latter also contributes the majority of the systematic error on the mass: $\pm 0.6$ MeV/$c^2$. These systematic errors are common to both $D^0$ decay modes. Averaging the two sets of values, we find that the $D^{*+}_{sJ}$ has a natural width of $16^{+5}_{-4} \pm 3$ MeV/$c^2$ and a mass of $2573.2^{+1.7}_{-1.6} \pm 0.8 \pm 0.5$ MeV/$c^2$, where the third error on the mass is due to the uncertainty in the $D^0$ mass.

Our width measurement implies that this new state decays strongly. Assuming this is so, angular momentum and parity conservation require that its spin-parity be in the so-called "normal" series: $0^+$, $1^-$, $2^+$, $3^-$, *etc.*

To measure the production cross section for $x \geq 0.7$ times the branching ratio to $D^0 K^+$, we have remeasured the yield in the second decay mode with the $x$ cut reduced to 0.7. We have also removed our cuts on $\cos\alpha_{K^+}$ in both modes, since the distribution of the signal events in this variable is unknown. For the $D^0 \to K^-\pi^+\pi^0$ mode, the number of events was reduced by 22% to account for the peaking of the background under the signal. An error of 9% was included in the systematic error to account for the uncertainty in this correction. Using the recent CLEO measurement of $\mathcal{B}(D^0 \to K^-\pi^+) = 3.91 \pm 0.19\%$ [8], the Particle Data Group's value for the ratio $\mathcal{B}(D^0 \to K^-\pi^+\pi^0)/\mathcal{B}(D^0 \to K^-\pi^+) = 3.10 \pm 0.26$ [5] and our measured luminosity of $2.16 \pm 0.02$ fb$^{-1}$, we find that for $x \geq 0.7$, the cross section times branching fraction is $\sigma(x \geq 0.7) \cdot \mathcal{B}(D^{*+}_{sJ} \to D^0 K^+) = 4.4 \pm 0.9 \pm 0.7$ pb. The first error reflects the uncertainty in the number of events, both statistical and systematic. It was calculated separately for the two modes and averaged. The second error is the systematic error common to both modes and is dominated by the uncertainty in the efficiencies.

We have also searched for the decay of this new resonance to $D^{*0}K^+$, $D^{*0} \to D^0\pi^0$. This mode is allowed for the $D^{*+}_{s2}$, but expected to be highly suppressed by the limited phase space. We used the same $D^0$ decay modes and similar cuts as in the previous analysis. In addition, we consider the helicity angle of the $\pi^0$ from the $D^{*0}$ decay. The helicity angle, $\theta_\pi$, is defined as the angle between the $D^{*+}_{sJ}$ and the $\pi^0$, both measured in the $D^{*0}$ rest-frame. For any meson with spin-parity in the "normal" series, the helicity angle of the $\pi^0$ from the $D^{*0}$ decay must have a $\sin^2\theta_\pi$ distribution. We require $|\cos\theta_\pi| \leq 0.75$.

We find no signal above background and set the following limit on the ratio of branching fractions:

$$\frac{\mathcal{B}(D^*_{sJ}(2573)^+ \to D^{*0}K^+)}{\mathcal{B}(D^*_{sJ}(2573)^+ \to D^0 K^+)} < 0.33$$

at the 90% confidence level. For the $D^{*+}_{s2}$, this ratio is predicted to be $\sim$ 0.1–0.16 [2].

In summary, we find a signal with a signifcance of more than six standard deviations for a meson decaying to $D^0K^+$. The mass and natural width of the new state are measured to be $2573.2^{+1.7}_{-1.6}\pm 0.8\pm 0.5$ MeV/$c^2$ and $16^{+5}_{-4}\pm 3$ MeV/$c^2$ respectively. Though its spin-parity is not established, it must be in the "normal" $0^+$, $1^-$, $2^+$, $3^-$ ... series. We tentatively identify this state as the $D_{s2}^{*+}$, the $2^+$ partner of the $D_{s1}(2536)^+$. It is $38.1^{+1.7}_{-1.6}\pm 0.8$ MeV/$c^2$ heavier than the $D_{s1}^+$, which has $J^P = 1^+$ [4]. This splitting is comparable to that seen between the $D_1(2420)^0$ and the $D_2^*(2460)^0$, the neutral $1^+$ and $2^+$ states. Thus the $D_{s1}^+$ and this new resonance appear to form a similar doublet. Its width is inconsistent with that predicted for a $0^+$ state, while both its width and decay modes are consistent with predictions for the $D_{s2}^{*+}$ [2].

We gratefully acknowledge the effort of the CESR staff in providing us with excellent luminosity and running conditions. This work was supported by the National Science Foundation, the U.S. Dept. of Energy, the Heisenberg Foundation, the SSC Fellowship program of TNRLC, and the A.P. Sloan Foundation.

# FIGURES

FIG. 1. $M^*$, "corrected" invariant mass, of $(K^-\pi^+[\pi^0])K^+$ combinations. Data points are for $K^-\pi^+[\pi^0]$ combinations in the $D^0$ signal region; the histogram shows $M^*$ for $(K^-\pi^+[\pi^0])K^+$ combinations where the $K^-\pi^+[\pi^0]$ combinations were chosen in $D^0$ sidebands.

FIG. 2. Histogram of $M^*(D^0K^+)$, with fit. The solid line shows the complete signal and background fitting functions. The sum of the background functions is shown by the dashed line. The dotted line shows just the polynomial used to represent the combinatoric background. The shape of the $D_J^*(2470)^+$ background function is shown at the bottom by the dash-dot line, with the area scaled up by a factor of 5.

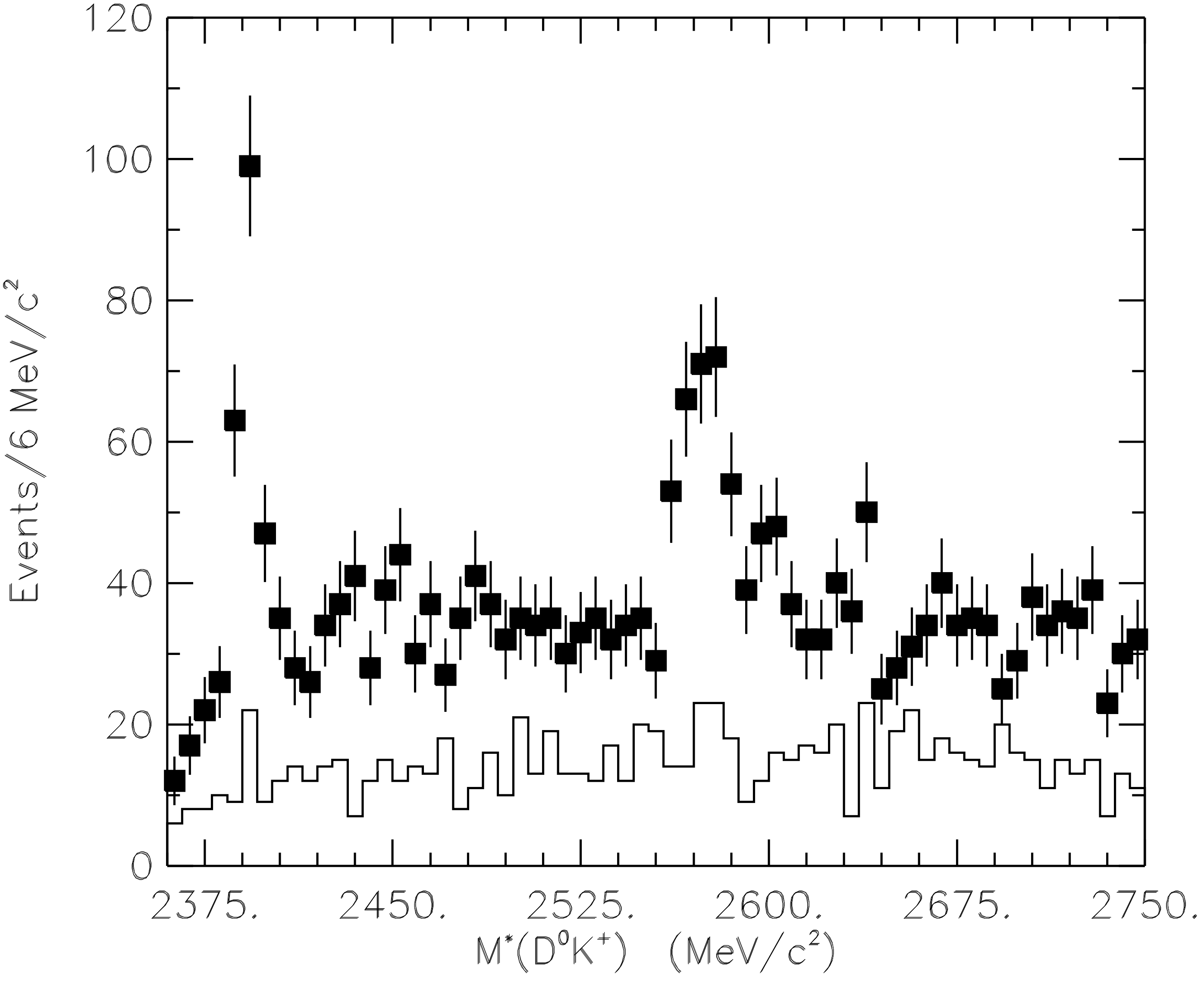

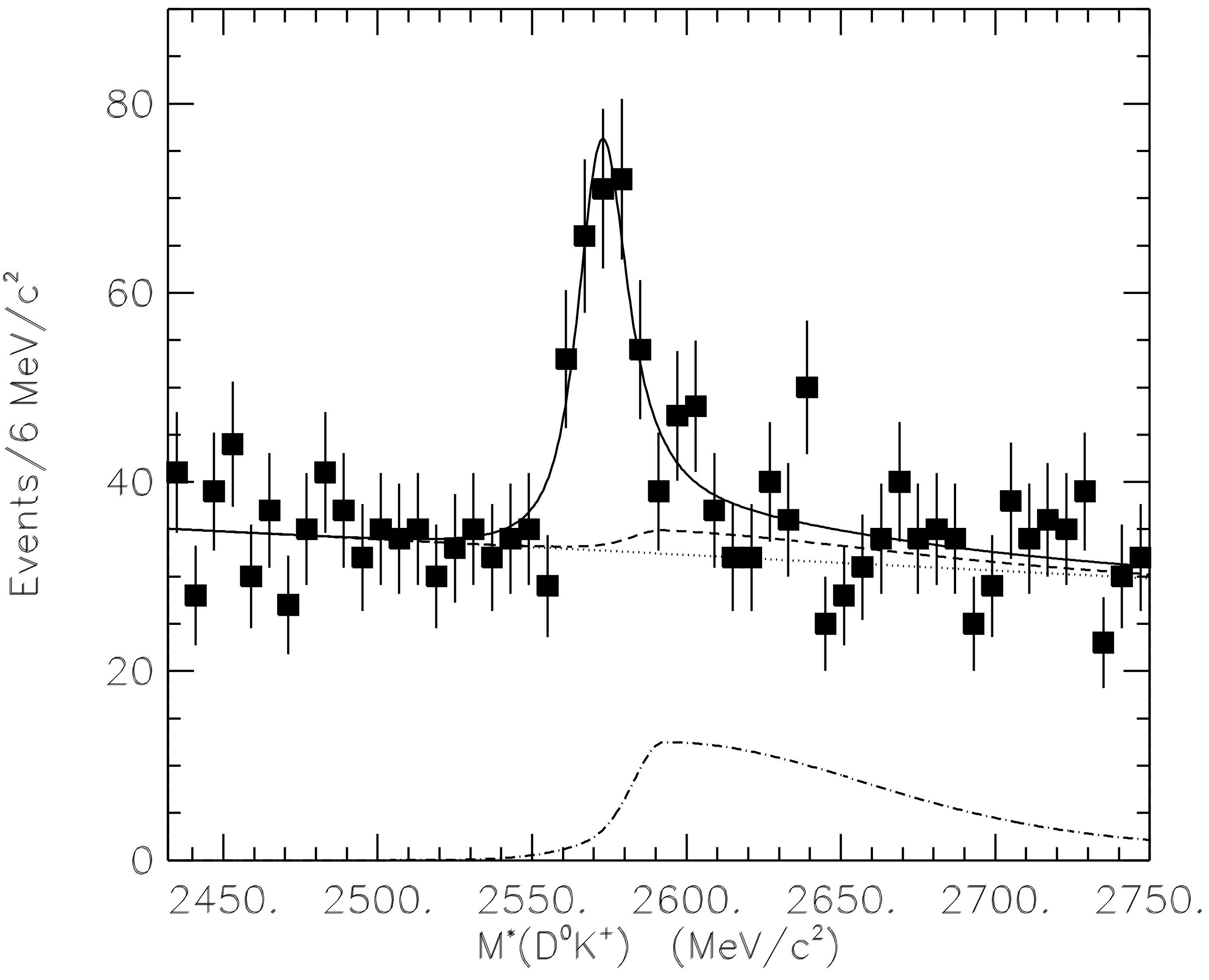